\documentclass[aps,prb,twocolumn,floatfix,groupedaddress,a4paper]{revtex4}

\usepackage{amsmath}
\usepackage{epsfig}

\begin{document}

\title{Fine structure and magneto-optics of exciton, trion, and charged biexciton states in single InAs quantum dots emitting at 1.3 $\mu$m}

\author{N.\ I.\ Cade}
\email{ncade@will.brl.ntt.co.jp}
\author{H.\ Gotoh}
\author{H.\ Kamada}
\author{H.\ Nakano}
\affiliation{NTT Basic Research Laboratories, NTT Corporation, Atsugi, 243-0198 Japan}

\author{H.\ Okamoto}
\affiliation{NTT Photonics Laboratories, NTT Corporation, Atsugi, 243-0198 Japan}

\begin{abstract}
We present a detailed investigation into the optical characteristics of individual InAs quantum dots (QDs) grown by metalorganic chemical
vapor deposition, with low temperature emission in the telecoms window around 1300 nm. Using micro-photoluminescence (PL) spectroscopy we
have identified neutral, positively charged, and negatively charged exciton and biexciton states. Temperature-dependent measurements reveal
dot-charging effects due to differences in carrier diffusivity. We observe a pronounced linearly polarized splitting of the neutral exciton
and biexciton lines ($\sim$250 $\mu$eV) resulting from asymmetry in the QD structure. This asymmetry also causes a mixing of the excited
trion states which is manifested in the fine structure and polarization of the charged biexciton emission; from this data we obtain values
for the ratio between the anisotropic and isotropic electron-hole exchange energies of $\tilde{\Delta}_{1}/\tilde{\Delta}_{0} \approx$
0.2--0.5. Magneto-PL spectroscopy has been used to investigate the diamagnetic response and Zeeman splitting of the various exciton
complexes. We find a significant variation in g-factor between the exciton, the positive biexciton, and the negative biexciton; this is
also attributed to anisotropy effects and the difference in lateral extent of the electron and hole wavefunctions.
\end{abstract}


\maketitle
\section{Introduction}
\label{intro}

Semiconductor quantum dots (QDs) have attracted considerable attention in recent years as they exhibit novel optical and electronic
phenomena,\cite{mowbray05} which increasing cannot be explained with an ``artificial atom'' type model.\cite{karrai04} High spatial
resolution spectroscopy can provide an detailed insight into the nature of the confinement potential of individual dots and the
quasiparticles that can form within them.\cite{zrenner00} The creation of charged exciton states (trions) in QDs is of particular interest
as these complexes are easily ionized in higher dimensional nanostructures. Over the last few years there have been many investigations
into the properties of QD trions in both II-VI\cite{turck01,patton03,akimov05} and III-V\cite{warburton00,moskalenko02,lomascolo02}
semiconductor materials; these studies have revealed a complex hierarchy of energies related to Coulomb, exchange, and correlation
interactions between the constituent electrons and holes.\cite{zunger04,rodt05}

The relative strengths of the isotropic and anisotropic parts of the electron-hole (\textit{eh}) exchange are manifested in the fine
structure splitting of the exciton state, which results from a reduction of the QD symmetry.\cite{kulakovskii99,bayer02b} This effect can
be studied in detail using magneto-optics;\cite{bayer00} however, to our knowledge, there have been no previous investigations into the
properties of magneto-excitons in individual long-wavelength ($>$1 $\mu$m) QDs. Furthermore, the specific growth conditions required to
produce these structures,\cite{seravalli05} and their enhanced quantum-confinement,\cite{liu05} are likely to have a significant effect on
the carrier interaction energies.  The development of semiconductor QD based devices for quantum optics and quantum information processing
necessitates a detailed understanding of the nature of these interactions: the trion state has no fine structure and is therefore suitable
for use as a single photon source,\cite{ulrich05} and the strength of the anisotropic \textit{eh} exchange energy is an important issue in
spin control systems.\cite{kroutvar04}

In previous studies the charge state of the QD has been varied using electrical injection of carriers\cite{warburton00,finley01b,bracker05}
or photodepletion effects.\cite{hartmann00,regelman01} However, with both techniques a comparative investigation of positive and negative
charged species is complicated by requisite changes in the environmental conditions: in the former case the electrical field induces an
emission Stark shift, and the latter case requires an increased carrier population. Here, we present the results of a comprehensive
investigation into the optical characteristics of individual InAs QDs emitting at 1.3 $\mu$m; in addition to the formation of an
exciton-biexciton system, we \textit{simultaneously} observe recombination from positive and negative trion and biexciton states. These
emission lines have relative intensities that are found to be highly sensitive to temperature due to diffusive dot-charging effects.
Polarized photoluminescence (PL) spectroscopy shows that asymmetry induced fine structure and state-mixing are present in the various
exciton complexes, and from a systematic study of different dots we have quantified the magnitudes of \textit{eh} exchange energies.
Finally we present a magneto-optical investigation of single QDs and we compare the diamagnetic shifts and Zeeman splittings of the neutral
and charged complexes.

\section{Experiment}
\label{structure}

The QDs investigated in this paper were fabricated by low-pressure metalorganic chemical vapor deposition (MOCVD) on a (100) GaAs
substrate, using the dots-in-well (DWELL) technique:\cite{lester99} a thin InAs dot layer ($\sim$ 1.7 monolayers) was embedded in a 5 nm
In$_{0.12}$Ga$_{0.88}$As(:Bi) quantum well (QW) and the whole DWELL heterostructure grown between GaAs barrier layers. More details of the
growth procedure and general optical characteristics are presented in Ref.\ \onlinecite{cade05}.

To obtain single dot spectroscopy, mesa structures were fabricated by electron-beam lithography and dry etching. Zero-field micro-PL was
taken from individual mesas excited by an Ar$^{+}$ laser (2.54 eV) focused to a $\sim$1 $\mu$m spot. Unless otherwise stated, the sample
temperature was maintained at 5 K in a continuous-flow He cryostat. Magneto-PL measurements were performed at 10 K with the field aligned
along the growth direction (Faraday geometry), using a diode laser (2.32 eV) focused to a $\sim$5 $\mu$m spot. The polarization of the
luminescence was analyzed using a linear polarizer and quarter-wave retarding plate. In all cases the luminescence was dispersed in a 0.5 m
spectrometer and detected with a nitrogen cooled InGaAs photodiode array ($1024\times1$).

\section{Results and discussion}
\label{results}

\subsection{Power dependence, peak assignment and thermal charging effects}
\label{powersection}

\begin{figure}[tb!]
\epsfig{file=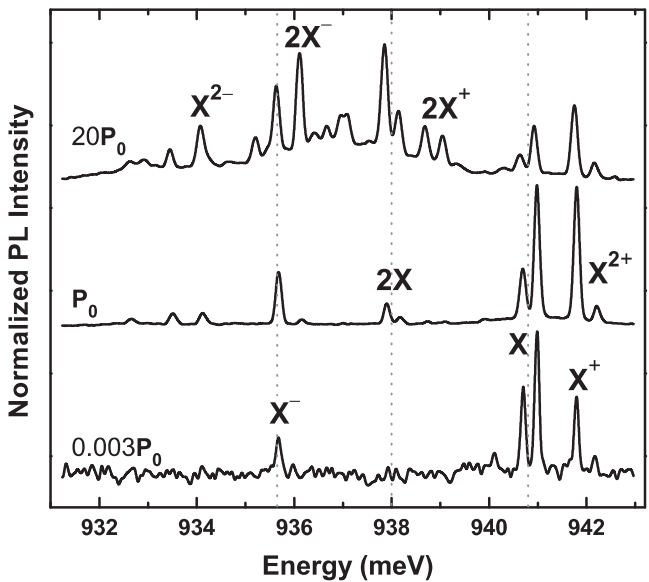,width=8.5cm} \caption{Normalized PL spectra from a single dot (QD1) at various excitation powers ($\textrm{P}_{0}=5$
W\,cm$^{-2}$). Peaks \textit{X} (2\textit{X}) and $X^{\pm}$, $X^{2\pm}$ (2$X^{\pm}$) are attributed to neutral and charged exciton
(biexciton) emission, respectively. } \label{power}
\end{figure}

PL spectra were taken for a 200 nm mesa over four orders of magnitude in excitation power, as shown in Fig.\ \ref{power}. At low powers the
spectra are composed of four narrow lines ($<$100 $\mu$eV, resolution limited); we have verified that all of these emission lines originate
from a \textit{single} dot in the mesa by comparing the PL spectra from many different mesas.\cite{cade05} These lines show a linear
increase in intensity over low excitation powers before saturating at $\sim$10 W\,cm$^{-2}$ (2P$_{0}$), and they are attributed to
recombination from the exciton \textit{X} and trion $X^{\pm}$ states. The exciton line shows a linearly polarized fine-structure splitting
of approximately 300 $\mu$eV; this will be discussed further in the next section.

With increasing power additional lines appear below the exciton energy. In particular, the 2\textit{X} doublet has a total intensity that
is a quadratic function of excitation power, and this is assigned to the biexciton state. The binding energy of $\sim$3 meV is consistent
with the values obtained by Kaiser \textit{et al}.\ \cite{kaiser02} for a similar strongly confined DWELL system. Other lines around
2\textit{X} also show a superlinear intensity behavior; these are assigned to charged biexciton states 2$X^{\pm}$ formed by the capture of
two \textit{eh} pairs into a charged dot. The lowest energy features in Fig.\ \ref{power} are attributed to multiply negatively charged
states ($X^{2-}$, $X^{3-}$ etc.), each of which is split into a multiplet through electron-electron and electron-hole exchange
interactions.\cite{urbaszek03,finley01b}

\begin{figure}[tb!]
\epsfig{file=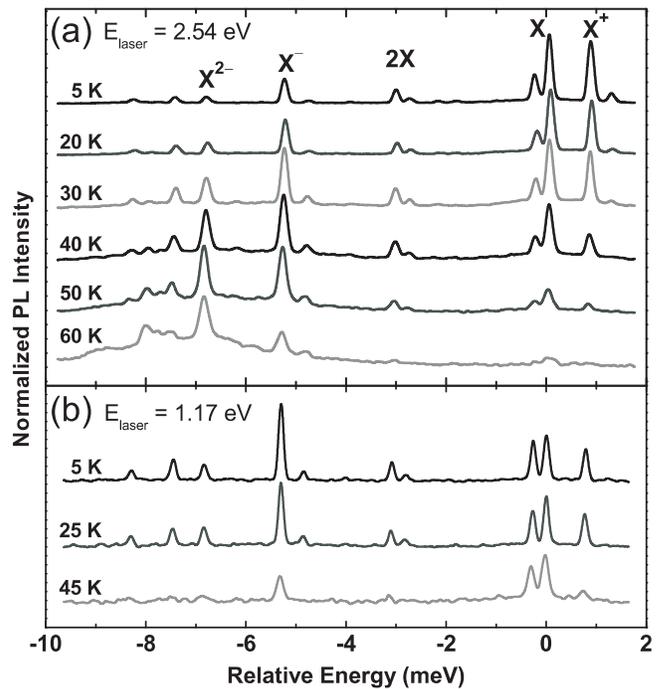,width=8.5cm} \caption{Normalized temperature-dependent PL spectra from QD1, with excitation (a) \textit{above} the
GaAs barrier and QW energy [2.54 eV, power P$_{0}$], (b) \textit{below} the QW energy [1.17 eV, 200P$_{0}$]. In both cases the spectra have
been shifted in energy to align the exciton \textit{X} doublet, and offset vertically.} \label{temp}
\end{figure}

Generally there are several possible origins for the excess carriers required for the creation of charged complexes: Nominally undoped GaAs
structures usually have a residual background doping leading to impurities in the vicinity of a dot. However, these impurities will result
in different emission spectra for each dot depending on the exact charge environment, whereas we obtain essentially identical spectra from
many dots separated by millimeters on the sample. There is also no evidence of photodepletion effects that are typically observed in doped
structures:\cite{hartmann00} the positive and negative trion species seen in Fig.\ \ref{power} are present for all excitation powers.

A second charging mechanism is the difference in carrier mobilities when \textit{eh} pairs are generated in the GaAs barrier by nonresonant
excitation. To investigate this effect PL spectra were taken with Ar$^{+}$ laser excitation (2.54 eV) from the same dot as Fig.\
\ref{power} (QD1), over a range of temperatures; this is shown in Fig.\ \ref{temp}(a). At 5 K the total integrated intensity of all the
emission lines of one charge type is approximately equal for both positive and negative species (at power P$_{0}$), which suggests that
there are random fluctuations in the excess charge within the dot. With increasing temperature there is a successive transfer of emission
intensity from the $X^{+}$ and \textit{X} lines to the $X^{-}$ and $X^{2-}$ lines; this arises from an increase in the electron diffusivity
relative to that of holes, effectively filling the dot with additional electrons.\cite{moskalenko02} This thermally enhanced diffusion
effect was confirmed by repeating the measurements using a YAG laser (1.17 eV) to excite below the GaAs barrier and InGaAs QW energy [Fig.\
\ref{temp}(b)]. In this case there is a monotonic decrease in intensity for all the trion lines with increasing temperature, due to the
absence of diffusive dot filling. Any charging originating from thermal activation of localized electrons should be insensitive to
excitation energy, which would lead to qualitatively similar behavior in both Figs.\ \ref{temp}(a) and \ref{temp}(b).

\subsection{Polarization properties and fine structure}
\label{finestructure}

\begin{figure}[tb!]
\epsfig{file=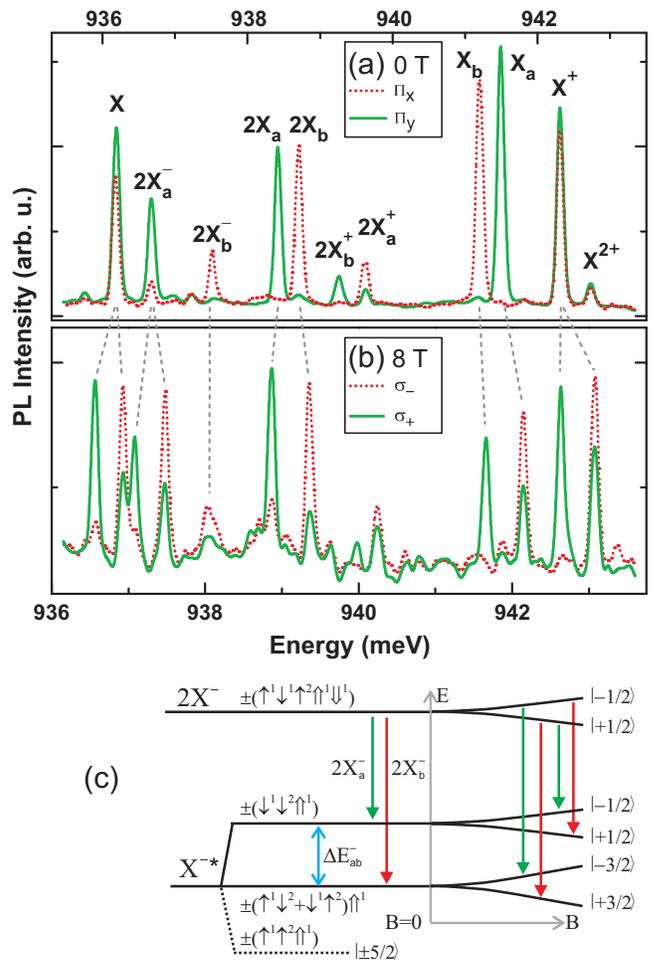,width=8.5cm} \caption{(Color online) (a) Linearly polarized PL components from QD1 at zero magnetic field. (b)
Circularly polarized PL components at $B=8$ T. (c) Energy level diagram showing allowed transitions from the charged biexciton ground state
to the trion triplet states, with and without a magnetic field applied in the growth direction. The spin projections of constituent
electrons $s_{z}= \pm 1/2$ and holes $j_{z}= \pm 3/2$ are denoted by $\pm (\uparrow^{l},\Uparrow^{l})$, respectively, for the
\textit{s}-shell ($l=1$) and \textit{p}-shell ($l=2$).  The 2$X^{+}$ has analogous transitions, except that the radiative $X^{+\ast}$
states are transposed.} \label{polar}
\end{figure}

The heavy-hole exciton in zinc blende based quantum dots (D$_{2d}$ point group) is fourfold degenerate and characterized by the angular
momentum components $M= s_{z}+j_{z}=\pm1, \pm2$, where $s_{z}= \pm 1/2$ is the electron spin and $j_{z}= \pm 3/2$ is the hole angular
momentum projection, respectively. The isotropic \textit{eh} exchange interaction splits this quartet into a radiative doublet $| \pm1
\rangle$ (bright excitons) and two nonradiative singlets comprising combinations of $| \pm2 \rangle$ states (dark excitons), with a
dark-bright exciton splitting $\Delta_{0}$. An in-plane anisotropy, caused by e.g.\ dot elongation and strain, will reduce the point group
symmetry; this results in an additional splitting of the radiative doublet into the states $X_{a,b}=\frac{1}{\sqrt{2}}(| +1 \rangle \mp |
-1 \rangle)$, separated by the anisotropic \textit{eh} exchange energy $\Delta_{1}$.\cite{bayer02b}

Figure \ref{polar}(a) shows polarized PL spectra from QD1, resolved along orthogonal axes using a rotatable polarizing prism. Both the
\textit{X} and the 2\textit{X} emissions consist of a doublet with the two components linearly polarized along the $\Pi_{\textrm{x}}$ and
$\Pi_{\textrm{y}}$ axes, which correspond closely to the (011) crystal axes. Both doublets have an identical splitting\cite{note1} and show
a mirror symmetry in the polarization sequence. These observations are consistent with recombination from the spin-singlet $| 0 \rangle$
biexciton ground state to the bright exciton states $X_{a}$ and $X_{b}$, with subsequent recombination to the $| 0 \rangle$ crystal ground
state.\cite{kulakovskii99,besombes00} Furthermore, with the application of a magnetic field \textit{B} there is a progressive evolution
towards circularly polarized emission [Fig.\ \ref{polar}(b)], as the mixed angular momentum states $X_{a,b}$ transform into pure $| \pm1
\rangle$ states.

In the trion ground state $X^{\pm}$ the two like-charges occupy the \textit{s}-shell with antiparallel spins and hence the exchange
interaction energies are quenched. Consequently the Kramers doublets are not split in the absence of a magnetic field even with a large dot
asymmetry, and only one line is seen for each trion in Fig.\ \ref{polar}(a).\cite{finley02} However, in the excited or ``hot'' trion state
$X^{\pm \ast}$ the additional charge resides in the \textit{p}-shell; the exchange energies are not quenched in this case and the
spin-triplet state is split into a set of three Kramers doublets through the \textit{eh} exchange interaction, as shown in Fig.\
\ref{polar}(c). The emission lines 2$X^{-}_{a}$ and 2$X^{-}_{b}$ result from transitions between the $| \pm\frac{1}{2} \rangle$ charged
biexciton ground state and the $| \mp\frac{1}{2} \rangle$ and $| \pm\frac{3}{2} \rangle$ $X^{- \ast}$ triplet states, respectively. The
total emission intensity ratio of $R_{0}=I_{b}/I_{a} \approx 0.5$ is consistent with a smaller transition probability for 2$X^{-}_{b}$, as
the final $| \pm\frac{3}{2} \rangle$ state has a nonradiative component.\cite{akimov05} Furthermore, the incomplete linear polarization of
these lines is direct evidence of state mixing caused by the anisotropic part of the \textit{eh} exchange;\cite{kavokin03} the larger
degree of polarization for 2$X^{-}_{b}$ as compared to 2$X^{-}_{a}$ is again a result of the difference in transition probabilities. A
similar analysis can be used to explain the characteristics of the 2$X^{+}_{a}$ and 2$X^{+}_{b}$ emissions; $R_{0}$ is also $\sim$0.5 for
these lines, but their relative intensity and degree of linear polarization have a mirror symmetry to the 2$X^{-}$ lines, as the radiative
triplet states are transposed in $X^{+\ast}$.

The exchange energies in the $X^{\pm \ast}$ states will be different to those discussed above for the neutral exciton, due to the
additional charge present in the \textit{p}-shell. Following the notation of Kavokin,\cite{kavokin03} the total exchange energies in
$X^{\pm \ast}$ are
\begin{equation}\label{exchange}
\tilde{\Delta}_{i}=\frac{1}{2}(\Delta_{i}^{1}+\Delta_{i}^{2})
\end{equation}
for $i=0$ (isotropic component) and 1 (anisotropic component); the superscripts 1 and 2 represent the exchange energy of the first and
second electrons (holes), respectively, with the hole (electron) in the $X^{- \ast}$ $(X^{+ \ast}$) state. For simplicity, we will continue
to write $\Delta_{1}\equiv \Delta_{1}^{1}$ unless explicitly required. The relative magnitudes of these exchange energies will be discussed
below.

\begin{figure}[tb!]
\epsfig{file=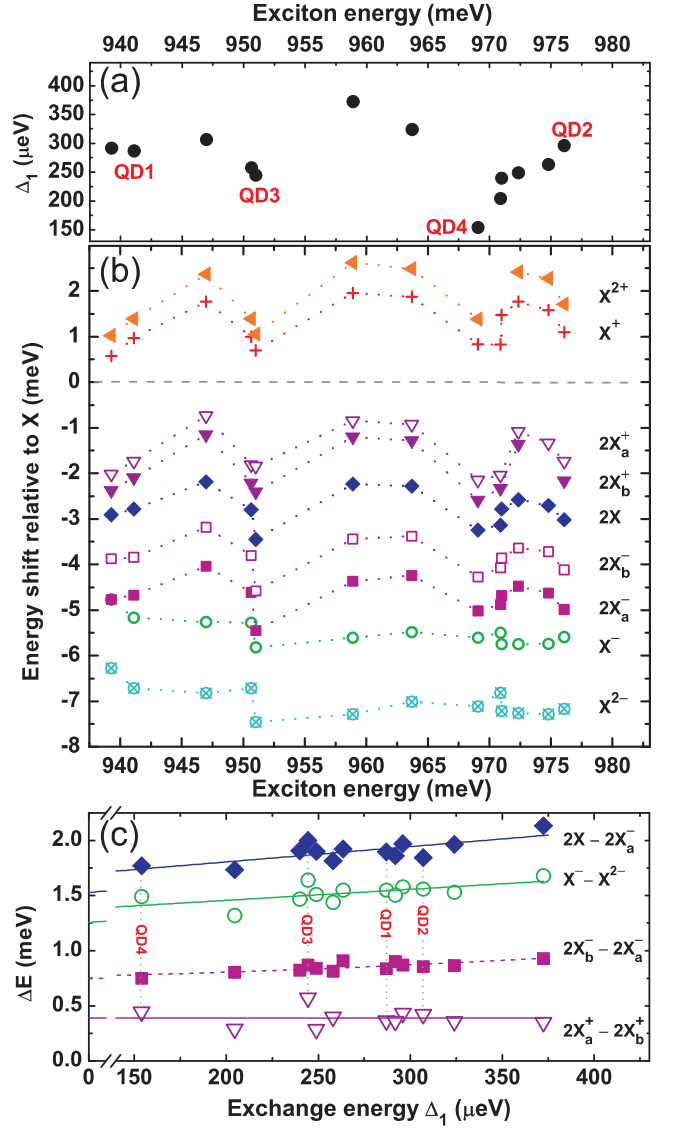,width=8.5cm} \caption{(Color online) Summary of the PL characteristics of 13 single dots; QD1--4 indicate the dots
referred to in the text: (a) Asymmetry induced fine-structure splitting $\Delta_{1}$ plotted against exciton emission energy. (b) Emission
energy of the exciton complexes relative to \textit{X}, plotted against exciton emission energy. (c) Energy difference $\Delta E$ between
emission lines in (b), as a function of $\Delta_{1}$ for each dot. Solid lines are linear data fits; the bottom line has a gradient of
zero. The dashed line is a fit using Eq.\ (\ref{anisofit}).} \label{survey}
\end{figure}

To gain a more detailed insight into the QD Coulomb energies, the emission spectra of 13 dots have been analysed and are summarized in
Fig.\ \ref{survey}. For each dot the value of $\Delta_{1}$ is shown in Fig.\ \ref{survey}(a) plotted against the exciton
energy.\cite{note1} The labels QD1--4 indicate the specific dots referred to throughout this report. There is no evidence of a direct
correlation between $\Delta_{1}$ and the \textit{X} emission energy; however, as $\Delta_{1}$ characterizes the asymmetry of the structure
it is highly sensitive to dot shape, especially in small dots.\cite{takagahara00}

For each dot, Fig.\ \ref{survey}(b) shows the energies of the other exciton complexes relative to the center of the \textit{X} doublet,
plotted against the \textit{X} emission energy. In all cases, the $X^{-}$ ($X^{+}$) line appears at a lower (higher) energy than the
exciton, which implies that the lateral extent of the single-particle wavefunction is smaller for the hole $l_{h}$ than for the electron
$l_{e}$.\cite{regelman01,warburton98} A semi-quantitative analysis of the binding energies of the exciton complexes has been performed: as
the quantization energy is large for these dots (giving a total \textit{s}-\textit{p} shell splitting of $\sim$90 meV),\cite{cade05} it is
possible to describe the charged excitons by treating the Coulomb interactions as perturbations to the single-particle states. Using the
exchange integrals calculated by Warburton \textit{et al}.\cite{warburton98} for a symmetric parabolic confinement potential, we find a
reasonable agreement with the observed $X^{\pm}$ binding energies for $l_{e}\approx 7.5$ nm and $l_{e}/l_{h}\approx 1.3$.  A full
comparison with theory will require a more detailed knowledge of the size, shape, and composition of the dots.

There are obvious similarities in the binding energy trends in Fig.\ \ref{survey}(b) for similar complexes. This is elucidated in Fig.\
\ref{survey}(c) which shows the energy separation $\Delta E$ between different emission lines plotted against $\Delta_{1}$ for each dot
surveyed. Despite the relatively large range in the overall emission energy, $\Delta E$ for a pair of lines varies by $<$300 $\mu$eV among
all the dots. This suggests that these QDs have a small size distribution, and the variations in \textit{X} emission energy are primarily a
result of fluctuations in the surrounding QW structure and process-induced strain effects.\cite{qiang94} For the negatively charged
complexes there is a direct correlation between $\Delta E$ and $\Delta_{1}$; in all cases the standard deviation from the solid best fit
line is $<$100 $\mu$eV. In particular, the splitting between the 2$X^{-}_{a,b}$ doublet, shown in Fig.\ \ref{polar}(c),
 is given by\cite{akimov05}
\begin{equation}\label{anisofit}
\Delta E_{ab}^{-}=\sqrt{(\tilde{\Delta}_{0})^{2}+2(\tilde{\Delta}_{1})^{2}};
\end{equation}
this gives a close fit to the data in Fig.\ \ref{survey}(c) (dashed line), where we have assumed that $\tilde{\Delta}_{0}$ is approximately
constant for these dots and $\tilde{\Delta}_{1}=\alpha \Delta_{1}$. We thence obtain values of $\tilde{\Delta}_{0}= 0.75$ meV and $\alpha =
1.05$, and from Eq.\ (\ref{exchange}) we find $\Delta_{1}^{2}/\Delta_{1}^{1} \approx 1.1$. This slight enhancement in the
2\textit{e}-1\textit{h} anisotropic exchange energy compared with that of 1\textit{e}-1\textit{h} is probably a result of the 2\textit{e}
\textit{p}-shell symmetry. For these dots the ratio between the anisotropic and isotropic \textit{eh} exchange energies thence has values
in the range $\tilde{\Delta}_{1}/\tilde{\Delta}_{0}$ = 0.21--0.52, which are similar to the values obtained for charged dots in II-VI
structures.\cite{akimov05}

In contrast to the 2$X^{-}$, the positively charged biexciton 2$X^{+}$ has a doublet splitting $\Delta E_{ab}^{+}$ that appears to be
insensitive to small variations in the dot asymmetry, as there is no obvious correlation with $\Delta_{1}$. This is consistent with the
smaller lateral extent of the hole wavefunction relative to that of the electron, as determined above. In general both $\tilde{\Delta}_{0}$
and $\tilde{\Delta}_{1}$ will differ between the 2$X^{-}$ and 2$X^{+}$, as the \textit{s}-\textit{p} exchange energies ($\Delta_{0,1}^{2}$)
between 1\textit{e}-2\textit{h} and 2\textit{e}-1\textit{h} are sensitive to the exact shape and overlap of the carrier wavefunctions. A
more detailed analysis would be facilitated by knowledge of the dark exciton energy levels using e.g.\ Voigt configuration
magneto-spectroscopy.\cite{bayer00}

\subsection{Magneto-photoluminescence}
\label{magnetoplsection}

\begin{figure}[tb!]
\epsfig{file=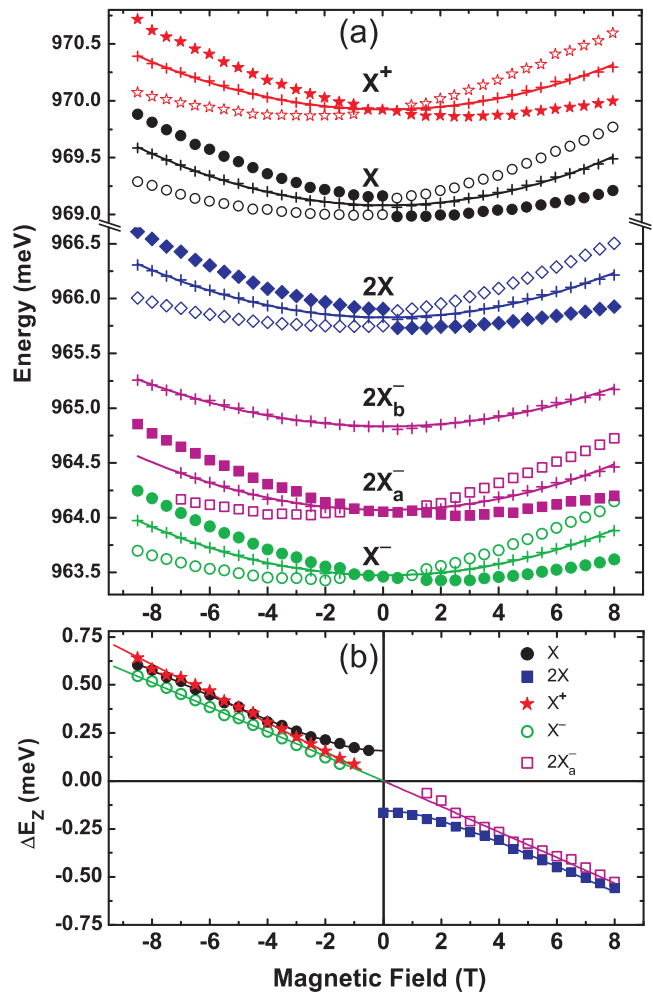,width=8.5cm} \caption{(Color online) (a) Magnetic field dependence of the PL emission lines from QD4. Open (closed)
symbols indicate preferentially $\sigma^{-}$ ($\sigma^{+}$) polarized emission. Crosses mark the center of each doublet and the solid
curves are quadratic fits. (b) Zeeman splitting $\Delta E_{Z}$ of the doublets in (a). The data for \textit{X} and 2\textit{X} have been
fit using Eq.\ (\ref{asymfit}). For clarity, only the negative (positive) range has been plotted for \textit{X} and $X^{\pm}$ (2\textit{X}
and 2$X^{-}$).} \label{magnetopl}
\end{figure}

The application of a magnetic field \textit{B} can reveal additional information about a dot's confinement potential and electronic states,
via the exciton diamagnetic response and spin-splitting. In the absence of fine structure the field-dependent exciton emission energy is
given by $E(B)=E_{0}+\gamma_{2} B^{2}\pm\frac{1}{2}g_\textrm{ex}\mu_{B}B$, where $E_{0}$ is the zero-field emission energy, $\gamma_{2}$ is
the diamagnetic coefficient, $g_\textrm{ex}$ is the effective g-factor of the exciton complex, and $\mu_{B}$ is the Bohr magneton. This
field-dependent splitting is shown in Fig.\ \ref{magnetopl}(a) for the neutral and charged complexes in QD4. The preferentially left
($\sigma^{-}$) and right ($\sigma^{+}$) circularly polarized components are plotted as open and closed symbols, respectively, and the
center position of each doublet is marked by a cross; a quadratic fit to these points is also shown by a solid line, from which
$\gamma_{2}$ has been obtained.

With the inclusion of the anisotropic \textit{eh} energy $\Delta_{1}$ the Zeeman splitting for the \textit{X} and 2\textit{X} states is
given by\cite{bayer99a}
\begin{equation}\label{asymfit}
 \Delta E_{Z} \equiv E(\sigma^{+})-E(\sigma^{-})=\sqrt{(g_\textrm{ex}\mu_{B}B)^{2}+(\Delta_{1})^{2}}
\end{equation}
where $E(\sigma^{\pm})$ is the energy of the $\sigma^{\pm}$ polarized emission. This gives an excellent fit to the data in Fig.\
\ref{magnetopl}(b) from which we obtain values of $g_{X}$ and $g_{2X}$. For the charged exciton complexes, $g_\textrm{ex}$ can be found
with a linear fit.

\begin{figure}[tb!]
\epsfig{file=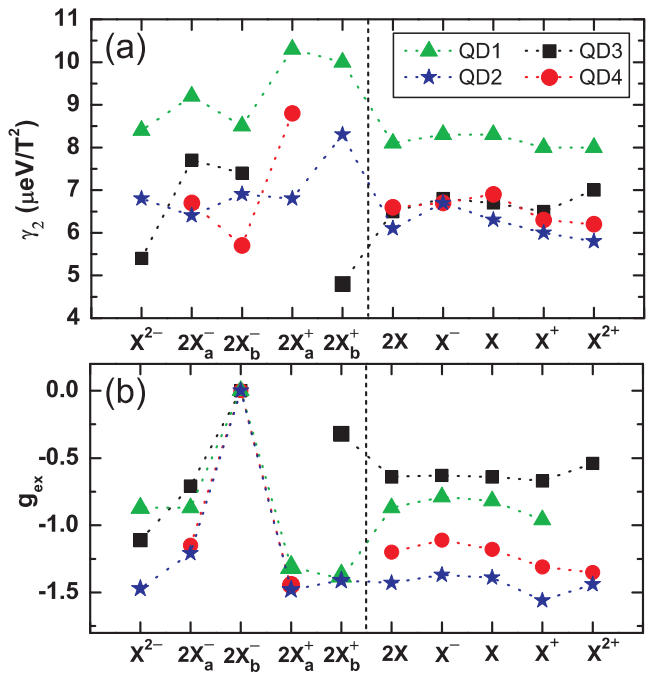,width=8.5cm} \caption{(Color online) Summary of (a) the diamagnetic coefficient $\gamma_{2}$, and (b) the g-factor
$g_\textrm{ex}$ of the exciton complexes, for QD1-4. Symbol size is greater than the absolute error in the data.} \label{gfactor}
\end{figure}

For excitons in strongly confined dots, where the quantization energy is larger than a typical Coulomb energy,  $\gamma_{2}$ is dependent
on the extent of the constituent carrier wavefunctions $l_{e}$ and $l_{h}$; the diamagnetic shift is given by the difference in the shifts
of the initial and final states, and will thus be identical for all exciton complexes with 1\textit{e}-1\textit{h} optical recombination.
Figure \ref{gfactor}(a) summarizes the values of $\gamma_{2}$ for the complexes seen in dots QD1--4. All four dots show a small diamagnetic
shift ($<$10 $\mu$eV\,T$^{-2}$) consistent with strong confinement; for each dot the neutral and charged states on the right of Fig.\
\ref{gfactor}(a) have equal values of $\gamma_{2}$ to within $\sim$0.5 $\mu$eV\,T$^{-2}$, as expected. The larger variations in
$\gamma_{2}$ for the charged biexciton complexes may originate from a small magnetic field dependence in the associated g-factor, which
will also introduce a quadratic \textit{B} term that is indistinguishable from the diamagnetic response.

The values of $g_\textrm{ex}$ shown in Fig.\ \ref{gfactor}(b) are consistent with the results of Nakaoka \textit{et al}.\cite{nakaoka05}
for InAs dots fabricated with a strain-reducing layer; the difference in $g_{X}$ among the dots originates from size, shape, and strain
variations. The degree of circular polarization seen in the spectra of each dot at 8 T is also consistent with the relative sizes of
$g_{X}\mu_{B}B$ and $\Delta_{1}$: QD1 and QD3 with smaller values of $|g_{X}|$ exhibit elliptically polarized emission even at 8 T, as seen
in Fig.\ \ref{polar}(a), whereas QD2 and QD4 emit almost completely circularly polarized light at $B=4.5$ T.

In an isolated dot the trion g-factor is predicted to be identical to that of the neutral exciton, as the excess carrier is present in the
initial and final states of the transition;\cite{schulhauser04} this is seen for QD3 in Fig.\ \ref{gfactor}(b). For the other dots studied
the variations in g-factor between \textit{X}, 2\textit{X}, and $X^{\pm}$, although small, are still greater than the experimental error
[as seen in Fig.\ \ref{magnetopl}(b)], and these variations show an identical trend in each case. The presence of dopants in the immediate
vicinity of a dot might lead to perturbation of the band structure and a significant deviation in g-factor between charged and neutral
excitons;\cite{bayer99b} however, the absence of any obvious emission from dark exciton states suggests that impurity related valence band
mixing effects are minimal for these dots,\cite{besombes00} as discussed in Sec.\ \ref{powersection}. It is likely that the variation
between $g_{2X^{-}_{a}}$ and $g_{X}$  is a result of state mixing caused by the anisotropic \textit{eh} exchange; furthermore, the
\textit{p}-shell electron may have a g-factor that is sensitive to the specific confinement conditions.\cite{akimov05} Mixing of the bright
$| \mp\frac{3}{2} \rangle$ and dark $| \pm\frac{5}{2} \rangle$ triplet states in the $X^{- \ast}$ could be responsible for the anomalous
zero spin-splitting of the 2$X^{-}_{b}$ line; this is unlikely to be a measurement error as it is observed for all four dots. However, a
more detailed analysis has not been possible due to the weak emission intensity of the 2$X^{-}_{b}$ line and the close proximity of other
spectral lines. In contrast to the 2$X^{-}$ lines, both 2$X^{+}$ emission lines show relatively large spin-splittings for three of the
dots, which are approximately equal ($g_{2X^{+}}\approx -1.45$) and independent of $g_{X}$. This is consistent with the data in Fig.\
\ref{magnetopl}(c) which shows that the $X^{+ \ast}$ state mixing is relatively insensitive to variations in the dot shape, due to the
smaller extent of the hole wavefunction.

\section{Summary}

We have analysed the PL spectra of single InAs QDs grown by MOCVD using a strain-relieving DWELL structure. In addition to the formation of
an exciton-biexciton system, we have simultaneously observed emission from both positive and negative trions and charged biexciton states
at 5 K. The relative intensities of these lines are highly sensitive to temperature and we find a thermal dot-charging effect which is
attributed to a difference in photogenerated carrier diffusivities. Analysis of the emission polarization reveals a large splitting of the
\textit{X} and 2\textit{X} lines ($\sim$250 $\mu$eV) due to the anisotropic electron-hole exchange interaction. The isotropic part of the
exchange interaction also splits the excited trion triplet states, and gives rise to a mirror symmetry in the intensity and polarization
degree of the charged biexciton emission doublets. From a study of the separation of the 2$X^{-}$ lines in different dots, we have
determined the ratio between the anisotropic and isotropic exchange energies. In contrast, the 2$X^{+}$ doublet splitting appears
uncorrelated with the degree of dot asymmetry; this is also manifested in the approximately constant g-factor for these lines, determined
by magneto-spectroscopy. These effects are likely to be the result of differences in the shape and extent of the carrier wavefunctions, and
warrant further investigation.

\section{Acknowledgements}

The authors are grateful to T.\ Tawara at NTT Basic Research Laboratories for etching the mesa structures, and S.\ Hughes for programming
assistance. One of the authors (N.I.C.) would like to thank A.\ Akimov for his helpful discussions. This work was partly supported by the
National Institute of Information and Communications Technology (NICT).

\end{document}